\documentclass[%
reprint,
superscriptaddress,
 amsmath,amssymb,
 aps,
pra,
longbibliography,
floatfix,
]{revtex4-2}
\usepackage{graphicx,subcaption}
\usepackage{color}
\usepackage{hyperref}
\usepackage{float}
\usepackage{array, multirow}
\usepackage{amsmath}
\usepackage{physics}
\usepackage{newtxtext}              
\usepackage{booktabs}
\usepackage{xcolor} 
\usepackage{soul}
\usepackage{url}
\usepackage{qcircuit}
\usepackage[margin=0.8in]{geometry} 

\hypersetup{colorlinks, 
    linkcolor={blue!75!black!80!yellow},
    citecolor={blue!75!black!80!yellow}, 
    urlcolor={blue!75!black!80!yellow}
    }

\begin{document}

\title{Singular value decomposition quantum algorithm for quantum biology}

\author{Emily K. Oh}
\affiliation{Department of Chemistry, Washington University in St. Louis, St. Louis, MO 61630 USA}

\author{Timothy J. Krogmeier}
\affiliation{Department of Chemistry, Washington University in St. Louis, St. Louis, MO 61630 USA}

\author{Anthony W. Schlimgen}
\affiliation{Department of Chemistry, Washington University in St. Louis, St. Louis, MO 61630 USA}

\author{Kade Head-Marsden}
\affiliation{Department of Chemistry, Washington University in St. Louis, St. Louis, MO 61630 USA}
\email{head-marsden@wustl.edu}


\begin{abstract}
There has been a recent interest in quantum algorithms for the modelling and prediction of non-unitary quantum dynamics using current quantum computers. The field of quantum biology is one area where these algorithms could prove to be useful, as biological systems are generally intractable to treat in their complete form, but amenable to an open quantum systems approach. Here we present the application of a recently developed singular value decomposition algorithm to two systems in quantum biology: excitonic energy transport through the Fenna-Matthews-Olson complex and the radical pair mechanism for avian navigation. We demonstrate that the singular value decomposition algorithm is capable of capturing accurate short- and long-time dynamics for these systems through implementation on a quantum simulator, and conclude that while the implementation of this algorithm is beyond the reach of current quantum computers, it has the potential to be an effective tool for the future study of systems relevant to quantum biology.
\end{abstract}

\maketitle


The majority of real physical systems interact with their environments in a non-trivial way. This is especially true for systems of biological relevance, where there is often a large and complex environment surrounding any energy or information transport process. Modelling these processes exactly is frequently computationally intractable; however, they are amenable to an open quantum system treatment.~\cite{Plenio2008} Standard methods in open quantum systems, such as the Lindblad equation,~\cite{Lindblad1976,Gorini1976, Breuer2002} are capable of accurately describing a variety of biologically relevant dynamical processes, including excitonic energy transport in photosynthetic light-harvesting antennae,~\cite{Plenio2008, Mohseni2008, Caruso2009, Chin2010, Caruso2010, Skochdopole2011, Mazziotti2012} radical pair mechanisms for avian navigation~\cite{Gauger2011, Carrillo2015, Stoneham2012} and other physiological functions,~\cite{Zadeh-Haghighi2022} and transport through ion channels.~\cite{Vaziri2010, Cifuentes2014, Bassereh2015, Jalalinejad2018}

An important aspect of recent quantum algorithm development has focused on the modelling of open quantum systems,~\cite{Miessen2023} which are systems that are not isolated but instead interact with their surroundings and are generally characterized by non-unitary dynamics. The challenge in developing gate-based quantum algorithms to capture these dynamical processes is that only unitary gates can be implemented on current quantum computers, but open quantum systems exhibit non-unitary time dynamics. A variety of algorithms have been developed to overcome this obstacle,~\cite{Garcia-Perez2020, Patsch2020,Kamakari2022} often using block encoding techniques.~\cite{Hu2020, HeadMarsden2021, Schlimgen2021, Hu2022, Schlimgen2022, Schlimgen2022PRR, Suri2023} Recently, two of the authors used classical computation of the singular value decomposition (SVD) of the time propagating operator, followed by an implementation of the dynamics with the singular value matrix on a quantum device.~\cite{Schlimgen2022} While this algorithm requires a non-negligible classical cost, the non-unitary component is mapped entirely to the diagonal singular-value matrix, and this sparsity can be leveraged when encoding the dynamics in a quantum circuit. The SVD has effectively been used to consider open quantum system evolution and general non-normalized state preparation.~\cite{Schlimgen2022} Here, we will use this algorithm on an IBM QASM simulator~\cite{Qiskit} to model the non-unitary dynamics of two systems in quantum biology: excitonic energy transport through a photosynthetic light-harvesting antenna, and the radical pair mechanism for avian navigation.
 
\begin{figure}[h!]
    \begin{subfigure}[t]{0.49\columnwidth}
        \includegraphics[width = \columnwidth, trim = 15cm 6cm 12cm 2cm,clip]{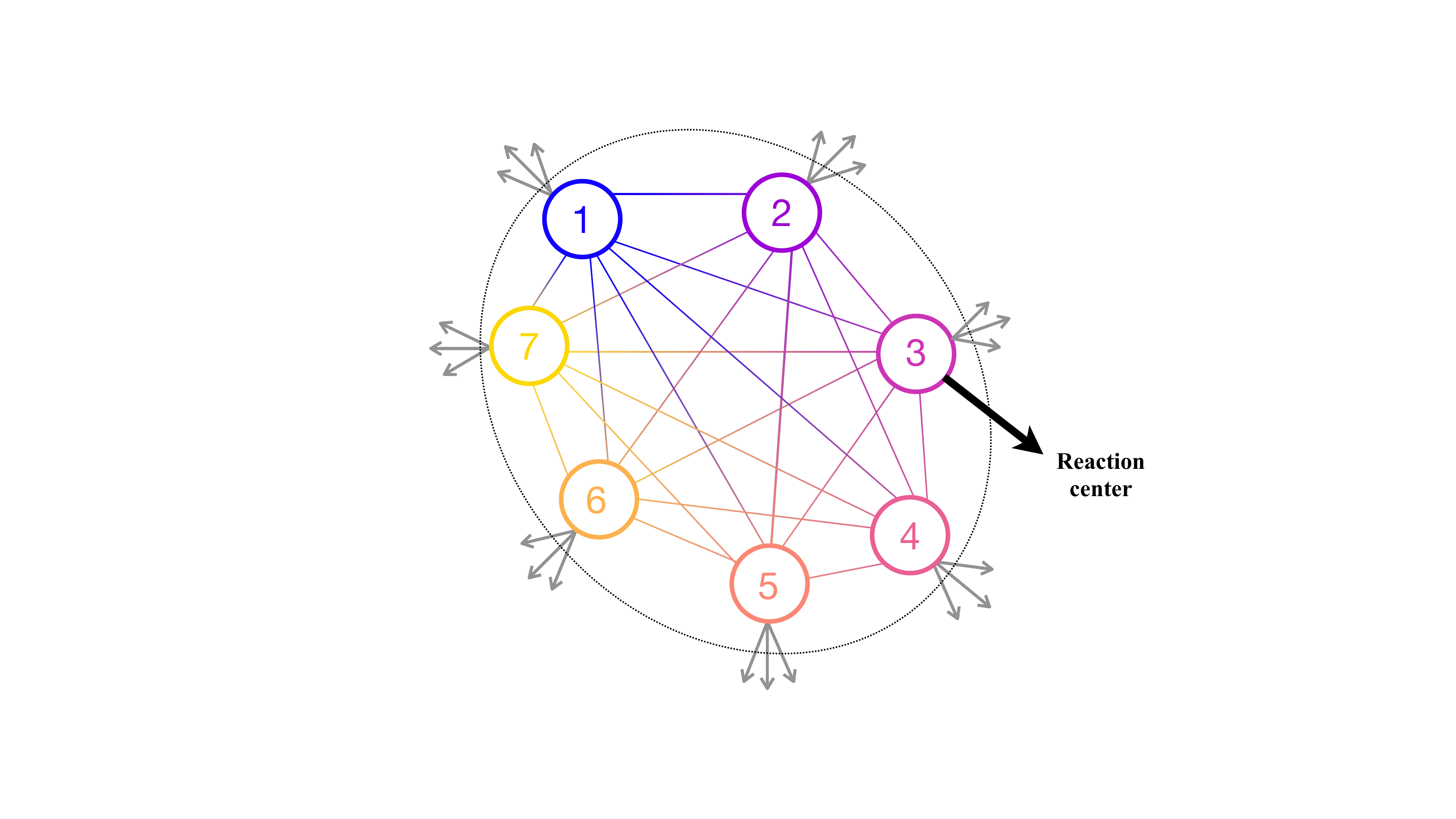}
        \caption{}
        \label{fig:FMO-schematic}
    \end{subfigure}
    \begin{subfigure}[t]{0.49\columnwidth}
        \includegraphics[width = \columnwidth, trim = 20.125cm 8cm 20.125cm 8cm, clip]{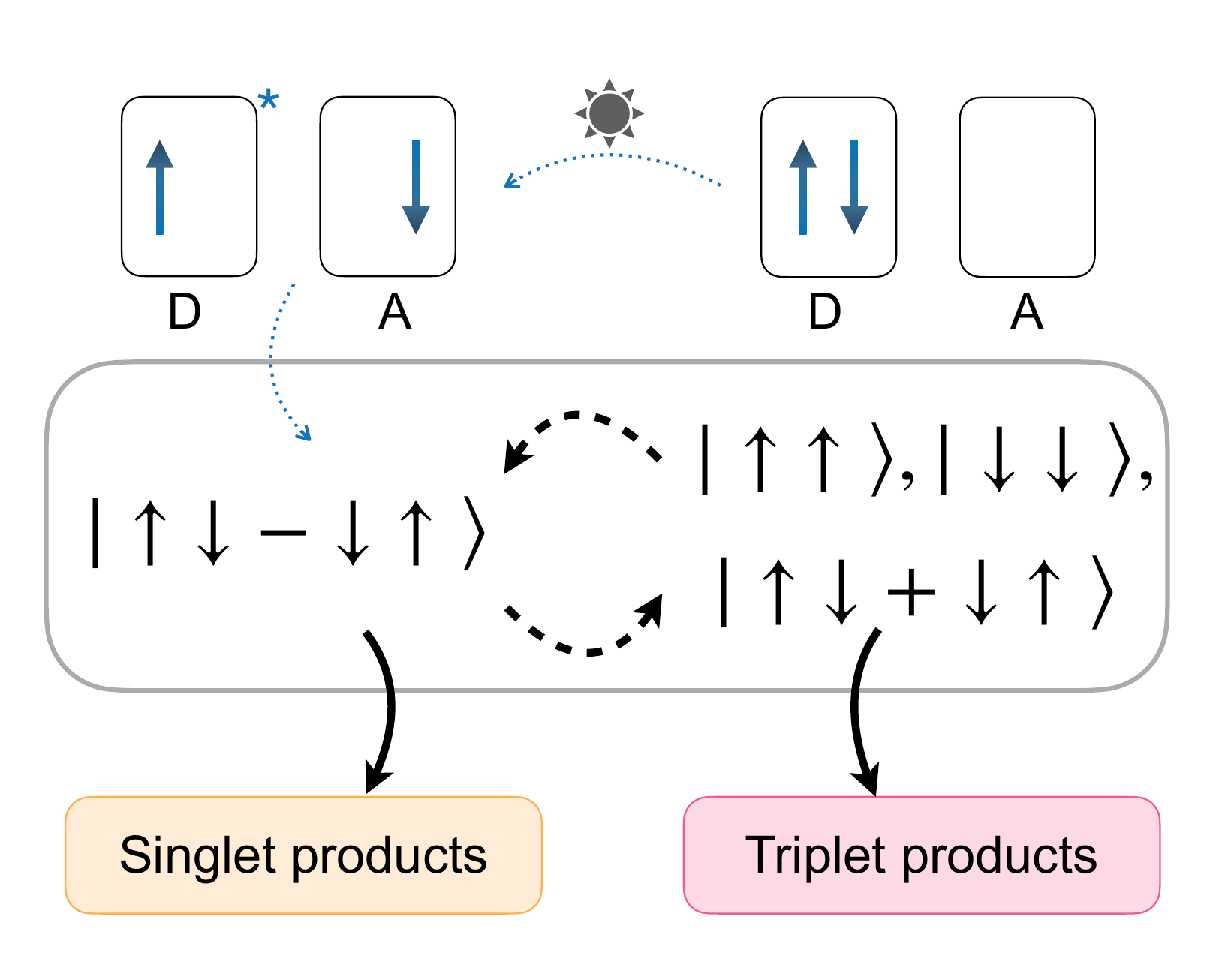}
        \caption{}
        \label{fig:RPM}
    \end{subfigure}
    \caption{(a) Schematic depiction of one trimer of the Fenna-Matthews-Olson complex, showing sites 1-7. The colors represent Hamiltonian terms, which include both on-site and between-site couplings. The grey arrows represent dissipation and decoherence due to the environment, and the black arrow represents the coupling of site 3 to the reaction center. (b) The radical pair mechanism: Excitation of the donor molecule, D, and the transfer of the electron to the acceptor, A, is shown along with the interconversion between singlet and triplet states and recombination. One of the electrons is coupled to a nuclear spin not shown in the above schematic.}
\end{figure}

First, we will consider the Fenna-Matthews-Olson (FMO) complex, which is a well-studied biological complex vital to photosynthetic light harvesting in green sulfur bacteria.~\cite{Blankenship2021} It exists as a trimer in the bacteria between the light-harvesting antenna and photosynthetic reaction center where it facilitates efficient exciton transfer. This is shown schematically in Figure~\ref{fig:FMO-schematic}, where an exciton is transferred into the complex on site 1, transported among the other sites, and eventually passes from site 3 to the reaction center where it can be converted into usable energy for the bacterium. While there have been extensive theoretical and experimental studies on this complex,~\cite{Fenna1975, Matthews1979, Olson2004, Lokstein2021, Shim2012, Cho2005} few have utilized quantum algorithms, and both the full 7 site system and the long-time dynamics have remained challenging to simulate.~\cite{Hu2022}

The second system that we will study is the radical pair mechanism (RPM) proposed for avian navigation.~\cite{Gauger2011} The RPM is theorized to explain how migratory birds can sense and navigate along Earth’s magnetic field.~\cite{Pauls2013, Schulten1978, Ritz2000, Zhang2014, Zhang2015} The basic scheme is represented in Figure~\ref{fig:RPM}. First, a donor molecule is excited by incoming light, causing the transfer of an electron from the donor to an acceptor molecule, creating a pair of coupled radicals. The radical pair is initially in the singlet state, but can interconvert between three triplet states as well. This conversion is partly determined by a coupled nuclear spin and the direction and strength of an external magnetic field. Depending on the spin state of the pair when they recombine, different chemical signals result. The yields of singlet and triplet products can therefore signal information about the orientation of the electron spin with respect to the external field. This particular application has also been extensively investigated theoretically,~\cite{Gauger2011,Pauls2013, Zhang2014, Zhang2015,Carrillo2015,Stoneham2012,Hore2016} including a quantum algorithm investigation,~\cite{Zhang2023} making it a good benchmark for the ability of a quantum algorithm to effectively capture a radical pair mechanism, which is prevalent in many other physiological processes.~\cite{Zadeh-Haghighi2022} 

First, we will review both the open quantum systems framework and the singular value decomposition algorithm. We will then present results using this algorithm for two systems, outlined above. Finally, we will discuss these results in the context of the potential for quantum algorithms to model and predict quantum systems of biological relevance.

\section*{Methods}

\subsection*{Lindblad approach to dissipative quantum systems}

A common model for the description of Markovian open quantum system dynamics is  the Gorini– Kossakowski–Sudarshan–Lindblad (GKSL) master equation,~\cite{Lindblad1976, Gorini1976, Breuer2002}
\begin{equation}\label{eq:lindbladME}
    \frac{d\rho}{dt} = -i[\hat{H},\rho] + \sum_i \gamma_i\big(\hat{C}_i\rho \hat{C}_i^{\dagger} - \frac{1}{2}\{ \hat{C
}_i^{\dagger}\hat{C}_i, \rho \}\big),
\end{equation}
where $\hat{H}$ is the system Hamiltonian, $\rho$ is the density matrix, and $\gamma_i$ are the decay rates corresponding to the physically relevant Lindbladian operators, $\hat{C}_i$. The first term represents the coherent evolution while the summation over Lindbladians represents the lossy, environmentally-drive dynamics. This equation can be written in a vectorized or unravelled master equation form, where Eq.~\ref{eq:lindbladME} is rewritten by reshaping the $r$ by $r$ density matrix into a vector of length $r^2$.~\cite{Havel2003, Schlimgen2022PRR} This can be done by stacking the columns of the original density matrix to produce a column vector, $\lvert \rho \rangle = \textrm{vec}(\rho)$. In this framework the Lindbladian superoperator is written as,
\begin{equation}\label{UME}
\begin{aligned}
    \hat{\mathcal{L}} &= -i\mathbb{I} \otimes \hat{H} + i\hat{H}^T \otimes \mathbb{I}\\
    &+ \sum_i \gamma_i(\hat{C}_i^* \otimes \hat{C}_i - \frac{1}{2} \mathbb{I} \otimes \hat{C}_i^{\dagger}\hat{C}_i - \frac{1}{2} \hat{C}_i^T\hat{C}_i^* \otimes \mathbb{I}),
\end{aligned}
\end{equation}
where $\mathbb{I}$ is the identity matrix and $*$, $\dagger$, and $T$ are the complex conjugate, adjoint, and transpose operations, respectively. The density matrix can then be propagated in time through,
\begin{equation}
    \ket{\rho(t)} = e^{\hat{\mathcal{L}}t}\ket{\rho(0)},
\end{equation}
where the propagation now occurs in Liouville space. 

\subsection*{Singular value decomposition based non-unitary quantum dynamics}

The Lindblad equation models non-unitary evolution, so the propagator $\hat{M} = e^{\hat{\mathcal{L}}t}$ needs to be mapped into a unitary form that can be implemented on current quantum devices. We begin with the SVD written as,
\begin{equation}
    \hat{M} = \hat{U}\hat{\Sigma}\hat{V}^\dagger
\end{equation}
where $\hat{U}$ and $\hat{V}^\dagger$ are unitary operators, and $\hat{\Sigma}$ is a real non-unitary diagonal operator. The diagonal operator can be dilated into a unitary (and diagonal) operator,
\begin{equation}
    \hat{U}_{\hat{\Sigma}} = \begin{pmatrix}
        \hat{\Sigma}_+ & 0 \\ 0 & \hat{\Sigma}_-
    \end{pmatrix}, 
\end{equation}
in which
\begin{equation}
    \hat{\Sigma}_{i,\pm} = \sigma_{i} \pm i\sqrt{\frac{1-\sigma_i^2}{\sigma_i^2}}\sigma_{i},
\end{equation}
where $\sigma_i$ are the singular values of $\hat{M}$.
 
\begin{figure}[ht]
    \centering
    \includegraphics[width = 0.8\columnwidth, trim = 5cm 21.5cm 9.5cm 4.5cm, clip]{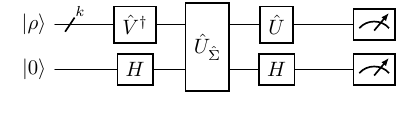}
    \caption{Quantum circuit for a non-unitary operator acting on a state, $\ket{\rho}$, by using the singular value decomposition and dilating the diagonal operator. The system spans $k$ qubits and the single additional ancilla bit is initialized in the ground state, $\ket{0}$.}
    \label{fig:SVDcircuit}
\end{figure}

Therefore, the non-unitary operator $\hat{M}$ can be implemented exactly on a quantum circuit as seen in Figure~\ref{fig:SVDcircuit}, where $k$ denotes that the system state spans multiple qubits. 
We compute the SVD of the exponential operator which yields a unique, but related, circuit for each time step. This circuit utilizes a linear combination of unitaries approach,~\cite{Childs2012} and results in a non-deterministic state which depends on the state of the ancilla qubit. When the ancilla is in state $\lvert 0 \rangle$, $\hat{M}$ is applied to the system qubit, $\hat{M}|\rho\rangle = \frac{1}{2}\hat{U}(\hat{\Sigma}_+ + \hat{\Sigma}_-)\hat{V}^\dagger |\rho \rangle$. When the ancilla qubit is in state $\lvert 1 \rangle$, then the procedure fails as $\frac{1}{2}\hat{U}(\hat{\Sigma}_+ - \hat{\Sigma}_-)\hat{V}^\dagger$ is applied to the system register. Notably, only one ancilla qubit is required, and the success probability does not depend on system size.
 
If a unitary matrix is size $r^2$, then it can be mapped to $k$-qubit unitary gates where $k \geq \log_2(r^2)$. The dilation of a non-unitary matrix adds an additional qubit resulting in $d = k+1$ qubits required to simulate the SVD of the non-unitary operator. The dilated singular value matrix can be implemented exactly with $\mathcal{O}(2^{d+1})$ gates, although polynomially scaling approximations are also available.~\cite{Welch:2014vj} The unitaries $\hat{U}$ and $\hat{V}^\dagger$ each require $\mathcal{O}((d-1)^22^{2d - 2})$ gates. The total gate complexity of the SVD algorithm is therefore $\mathcal{O}(d^22^{2d-1})$.~\cite{Schlimgen2022} In the limit of large system size, the application of $\hat{U}$ and $\hat{V}^{\dagger}$ to the quantum register generates the most overhead. In the asymptotic limit, given access to the SVD, a $d$-qubit non-unitary operator can be applied to a quantum register for approximately twice the cost of a $d$-qubit unitary operator.

Beyond the gate complexity of the circuit, there are other cost factors to consider. First, the singular value decomposition is computed classically with a complexity $\mathcal{O}(r^3)$ where $r$ is the size of the decomposed operator. When computed numerically, the SVD scaling is prohibitive for arbitrarily large or complex matrices; however, operators used in the context of Noisy-Intermediate-Scale Quantum (NISQ) devices are modestly sized and the SVD is easily computed classically. In addition, physical processes may have SVDs which can be written analytically.~\cite{Schlimgen2022} Looking forward to the fault-tolerant regime, this classical cost could be avoided by utilizing a quantum algorithm to calculate the SVD.~\cite{Suri2023,Bravo-Prieto:2020aa} Secondly, to obtain accurate long-time dynamics, the unravelled or vectorized master equation must be used. This involves transitioning from a Hilbert space of size $r$ to the Liouville space of size $r^2$ which also spans a larger qubit space. This mapping requires a larger number of qubits and therefore an increase in complexity; however, it allows for simulating long-time dynamics without approximating the solution to the differential Lindblad equation.

\section*{Results}

\subsection*{Light-harvesting antennae}

The exciton dynamics in the FMO complex have been successfully modelled classically by the Lindblad equation,~\cite{Plenio2008, Mohseni2008, Caruso2009, Chin2010, Caruso2010, Skochdopole2011, Mazziotti2012,Hu2022} where the coherent or unitary components are described by the Hamiltonian,
\begin{equation}
    \hat{H}_{FMO} = \sum_{i=1}^7 \omega_i\sigma_i^+\sigma_i^- + \sum_{j\neq i} J_{ij}(\sigma_i^+\sigma_j^- + \sigma_j^+\sigma_i^-),
\end{equation}
where $\sigma_i^+$ and $\sigma_i^-$ are the creation and annihilation operators respectively, $\omega_i$ is the on-site coupling, and $J_{ij}$ is the coupling between sites $i$ and $j$. We use the coupling parameters from Ref.~\citenum{Plenio2008}, and the full Hamiltonian in matrix form can be found in Supplementary Eq. S2. 

In the schematic of the full system in Fig.~\ref{fig:FMO-schematic}, the Hamiltonian terms accounting for the on-site chromophore energies are depicted by circled numbers and their couplings by lines. The Lindbladian terms account for the transfer of the exciton from the third chromophore to the sink, which models the reaction center, as well as dephasing and dissipation to the ground state. Transfer to the sink is represented by black arrows, and dephasing and dissipation by grey arrows in the schematic. These Lindbladians take the form,
\begin{equation}
\begin{aligned}
    &\hat{C}_{deph}(i) = \sqrt{\gamma_{deph}}\ket{i}\bra{i}, &
    &\hat{C}_{diss}(i) = \sqrt{\gamma_{diss}}\ket{0}\bra{i}, \\
    &\hat{C}_{sink} = \sqrt{\gamma_{sink}}\ket{8}\bra{3},
\end{aligned}
\end{equation}
where $i$ is an integer on the range $[1,7]$, states $\ket{0}$ and $\ket{8}$ model the ground and sink sites, respectively, and $\gamma_{deph}$, $\gamma_{diss}$, and $\gamma_{sink}$ represent the corresponding rates of dephasing, dissipation, and transfer to the sink for the 7-site model. Previous work has focused on the dynamics of a subsystem of this complex,~\cite{Skochdopole2011, Hu2022} which includes only the first 3 chromophores. For this 3-site model, $i$ is an integer only on the range $[1,3]$, and the sink is given by state $\lvert 4 \rangle$ instead of $\lvert 8 \rangle$. For both models, the system is initialized with the excitation on site 1. All relevant parameters can be found in Supplementary Table S1.

\begin{figure}[ht!]
    \centering
    \begin{subfigure}[t]{0.79\columnwidth}
        \includegraphics[width = \columnwidth]{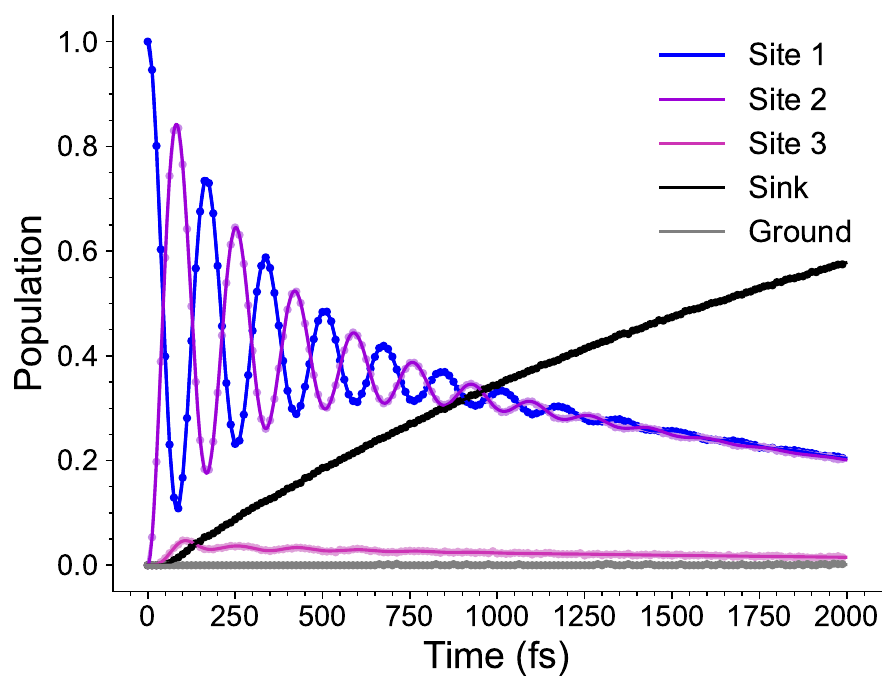}
        \caption{}
        \label{fig:FMO-3-site}
    \end{subfigure}
    \hfill
    \begin{subfigure}[t]{0.79\columnwidth}
        \includegraphics[width = \columnwidth]{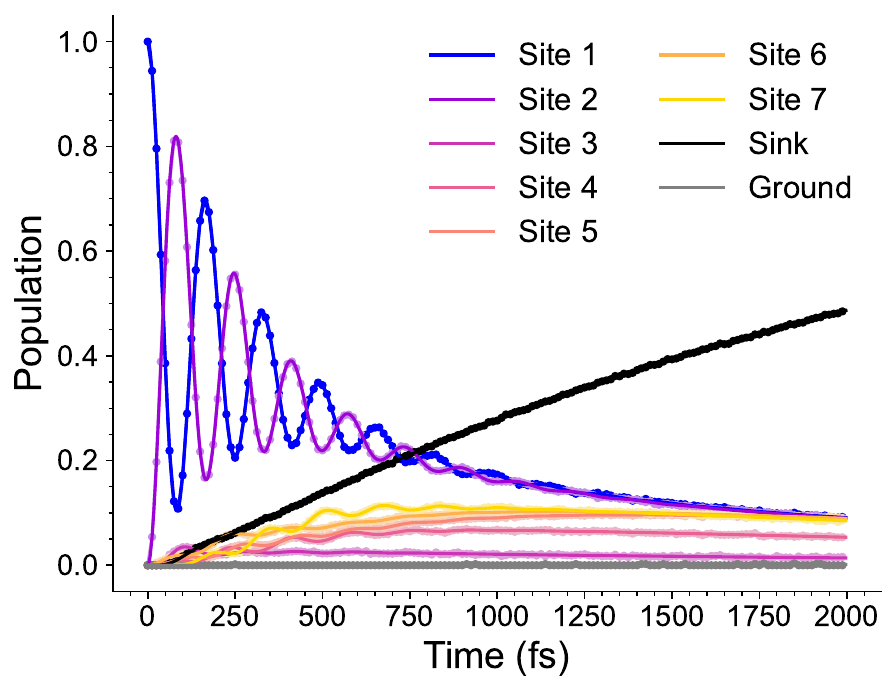}
        \caption{}
        \label{fig:FMO-results}
    \end{subfigure}
    \caption{Modeling the time evolution of the FMO complex. In (a), the results from the 3-site model are shown, and (b) shows the 7-site model results. For both plots, the lines indicate classical results and the dots the IBM QASM quantum simulation. A time step of $\delta t = 5$ femtoseconds (fs) and end time of 2000 fs were used. For each quantum measurement, $2^{19}$ samples were taken.}
    \label{fig:FMO}
\end{figure}

Utilizing the above parameters with the unravelled master equation in Eq.~\ref{UME} and performing the singular value decomposition on the resulting operator, $\hat{M} = e^{\hat{\mathcal{L}t}}$, we can obtain results for the 3-site model. The classical baseline and IBM QASM quantum simulation for the dynamics can be seen in Figure~\ref{fig:FMO-3-site} where the classical results are shown as solid lines and the quantum simulation as dots. A total duration of 2000 femtoseconds (fs) was used with a time step of 5 fs, utilizing a total of 6 qubits for the quantum simulation. For all simulation data collected, $2^{19}$ samples were used for consistency between trials and systems. These results show agreement between the quantum simulation and classical result for the entirety of the 2000 fs process, significantly extending the previous simulation time range while still maintaining accuracy.~\cite{Hu2022}

We also expand the focus to the entire 7 chromophore system dynamics, which becomes a 9 level system when a sink and a ground state are included. This is demonstrated in Figure~\ref{fig:FMO-results}, where again the classical solution is shown by solid lines and the results of IBM QASM quantum simulation using 8 qubits and $2^{19}$ shots are shown as dots. These quantum simulation results are also in excellent agreement with the classical solution.

Both the 3-site and 7-site models of excitonic dynamics in the FMO antenna demonstrate the capacity of the singular value decomposition algorithm to capture accurate dynamics on an IBM QASM quantum simulator, regardless of the length of time of the simulation. 

\subsection*{Avian compass}

The radical pair mechanism in the avian compass relies on the interconversion between singlet and triplet electronic states in an external magnetic field coupled to a single nuclear spin. This can be modelled with the Hamiltonian that takes the Zeeman and hyperfine interactions into account,~\cite{Gauger2011} 
\begin{equation}
    H = \hat{I} \cdot A \cdot \hat{S_1} + \gamma B \cdot (\hat{S_1} + \hat{S_2}),
\end{equation}
where $\hat{I}$ is the single nuclear spin operator, $A$ is the hyperfine tensor describing the anisotropic coupling between the nucleus and the first electron, $\hat{S}_{j}$ are the electron spin operators for electrons $j=1,2$, $\gamma$ is the gyromagnetic ratio, and $B$ is the applied magnetic field given by $B = B_0(\cos{\phi}\sin{\theta},\sin{\phi}\sin{\theta},\cos{\theta})$. The angles $\phi$ and $\theta$ describe the radical pair's orientation with respect to the external applied field, and based on symmetry $\phi$ can be set to zero. Due to the spatial separation of the electrons, only one electron is coupled to the nuclear spin in the Hamiltonian. The second electron is farther from, and thus much more weakly coupled to, the nuclear spin.~\cite{Gauger2011}

The singlet and triplet states in the electronic system can be written as,
\begin{equation}
\begin{aligned}
    &\ket{s} = \frac{1}{\sqrt{2}}(\ket{\uparrow}\otimes\ket{\downarrow} - \ket{\downarrow}\otimes\ket{\uparrow}) \\
    &\ket{t_-} = \ket{\downarrow}\otimes\ket{\downarrow} \\
    &\ket{t_0} = \frac{1}{\sqrt{2}}(\ket{\uparrow}\otimes\ket{\downarrow} + \ket{\downarrow}\otimes\ket{\uparrow}) \\
    &\ket{t_+} = \ket{\uparrow}\otimes\ket{\uparrow}, 
\end{aligned}
\end{equation}
where up and down arrows are used to represent $\alpha$ and $\beta$ spin states.

Coupling the electronic states with a single nuclear spin produces an 8 site model.  Shelving states $\ket{S}$ and $\ket{T}$ are added to indicate the yields of recombination products from the given radical pair conditions. They are only connected to the system through the following Lindblad operators,
\begin{equation}\label{RPMlindblad}
\begin{aligned}
    C_1 &= \ket{S}\bra{\uparrow,s}  &   C_2 &= \ket{T}\bra{\uparrow,t_0} &
    C_3 &= \ket{T}\bra{\uparrow,t_+} \\ 
    C_4 &= \ket{T}\bra{\uparrow,t_-} &
    C_5 &= \ket{S}\bra{\downarrow,s}  &   C_6 &= \ket{T}\bra{\downarrow,t_0} \\
    C_7 &= \ket{T}\bra{\downarrow,t_+}  &   C_8 &= \ket{T}\bra{\downarrow,t_-},
\end{aligned}
\end{equation}
where the $s$, $t_0$, $t_+$, and $t_-$ indicate the spin configuration of the radical pair of electrons, and the arrows signify the direction of the nuclear spin. The decay rates for all the shelving Lindbladians are, for the sake of simplicity, made equal and given by, $\gamma_{shelf}$. Operators $C_1$ and $C_5$ show recombination from the singlet radical configuration resulting in singlet products regardless of nuclear spin. The other six operators populate the triplet yield from the three possible triplet configurations for both nuclear spins.  

\begin{figure}[ht!]
    \centering
    \begin{subfigure}[t]{0.79\columnwidth}
        \includegraphics[width = \columnwidth]{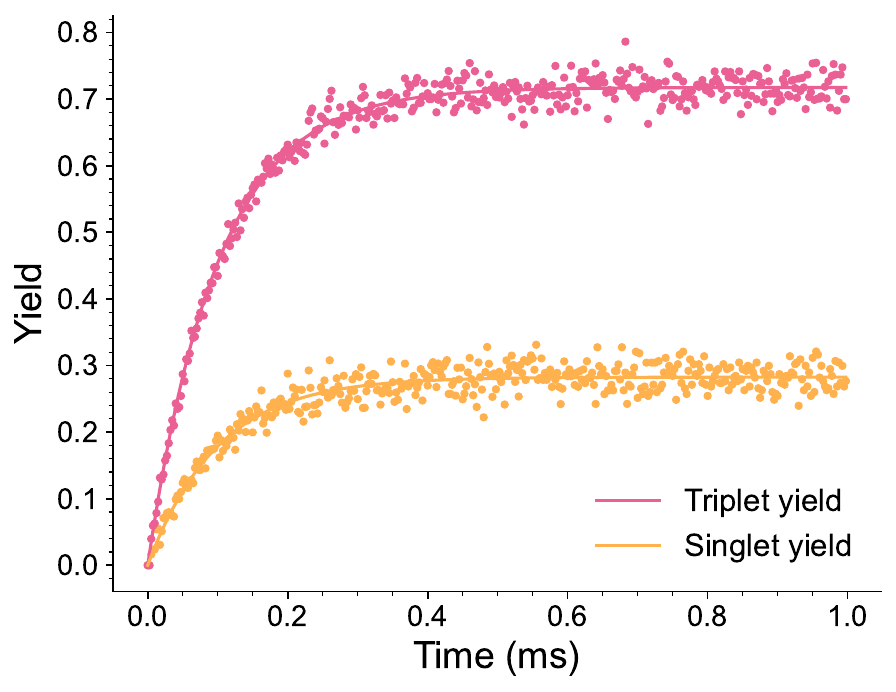}
        \caption{}
        \label{fig:RPM-yields}
    \end{subfigure}
    \hfill
    \begin{subfigure}[t]{0.79\columnwidth}
        \includegraphics[width = \columnwidth]{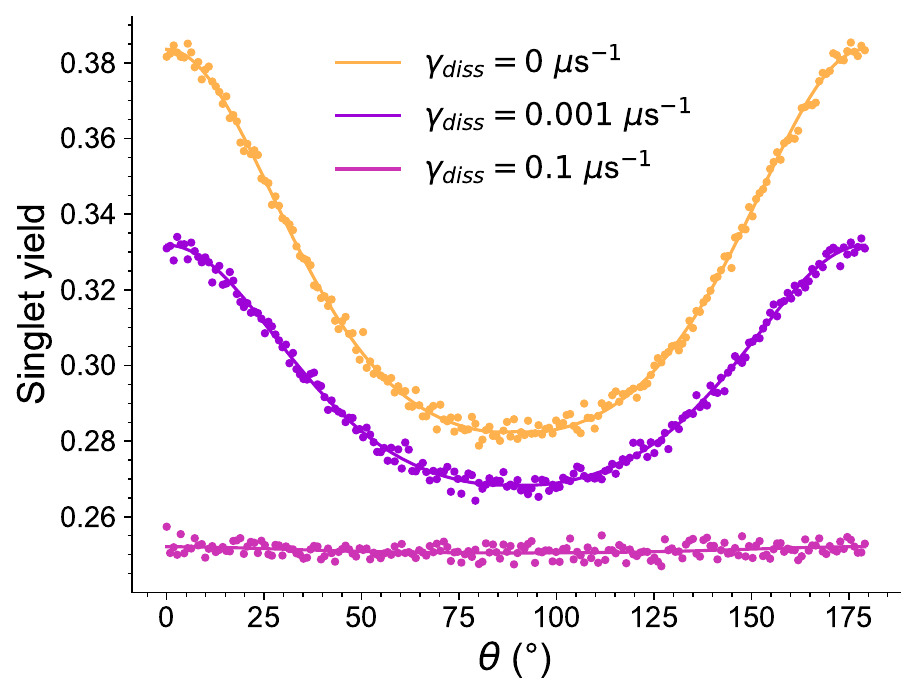}
        \caption{}
        \label{fig:RPM-dissipator}
    \end{subfigure}
    \caption{Modeling the radical pair mechanism for avian navigation. (a) Time evolution of singlet and triplet yields for the avian compass. Results are shown for an end time of 1 ms and a time step of $\delta t = 1.75 \cdot 10^{-3}$ ms. The angle between the radical pair and external field was fixed $\theta = \frac{\pi}{2}$. (b) Angle dependence of the avian compass singlet yields with and without noise from the environment. A theta jump of $\delta \theta = 0.9^\circ$ was used. For both plots, the smooth curves show classical results and the dots, IBM QASM quantum results. The rate of decay to the shelving states and quantum sampling were also set to $\gamma_{shelf} = 10^4$ and $2^{19}$, respectively.}
    \label{fig:RPM-results}
\end{figure}

This model was implemented through classical Lindbladian evolution and simulated through the SVD algorithm to find the time evolution of the singlet and triplet yields. An initial pure singlet and mixed nuclear state was used. The populations obtained from setting the external magnetic field to $B_0 = 47 \mu$T, the decay constant to $\gamma_{shelf} = 10^4$, and the angle to $\theta = \frac{\pi}{2}$ can be found in Figure~\ref{fig:RPM-yields}.  An orientation angle of $\theta = \frac{\pi}{2}$ indicates the external field is perpendicular to the radical pair. For the IBM QASM quantum simulation results, 8 qubits were required and $2^{19}$ samples were used. All relevant parameters are also documented in Supplementary Table S2.

This model so far assumes that there is no dissipation from the singlet or triplet electronic states, when in reality these states will also be dephasing while the radical pair is converting between them. This can be accounted for in the model with the addition of the following Lindbladians,
\begin{equation}
    \begin{aligned}
        C_9 &= \mathbb{I} \otimes \sigma_x \otimes \mathbb{I} & C_{10} &= \mathbb{I} \otimes \mathbb{I} \otimes \sigma_x\\
        C_{11} &= \mathbb{I} \otimes \sigma_y \otimes \mathbb{I} & C_{12} &= \mathbb{I} \otimes \mathbb{I} \otimes \sigma_y\\
        C_{13} &= \mathbb{I} \otimes \sigma_z \otimes \mathbb{I} & C_{14} &= \mathbb{I} \otimes \mathbb{I} \otimes \sigma_z,
    \end{aligned}
    \label{RPMDissipators}
\end{equation}
where $\sigma_i$ are the Pauli operators and $\mathbb{I}$ is the identity matrix. The Lindbladians in Eq.~\ref{RPMDissipators} use the decay constant $\gamma_{diss}$ and are padded with zeros to match the dimensionality of the shelving states. Considering three different decay rates, the singlet yields compared to the orientation angle between the radical pair and external magnetic field are shown in Figure~\ref{fig:RPM-dissipator}, where the classical solution is shown with solid lines and the IBM QASM quantum simulation is shown as dots. Again, 8 qubits were required and $2^{19}$ measurements were used for sampling, with the relevant parameters listed in Supplementary Table S2.

The algorithm results are in good agreement with the classical solution and demonstrates that greater dissipation rates lead to less differentiation in singlet and triplet yields across a range of orientation angles. Thus, the efficacy of the avian compass is suppressed with increased dissipation. For both the dissipation-free and dissipation models of the RPM, the SVD algorithm accurately captures the dynamics in all tested parameter regimes.

\section*{Discussion and Conclusions}

Here, we demonstrate the success of the SVD-based algorithm in capturing accurate long-time dynamics in two systems pertinent to quantum biology. The two systems we consider are the excitonic energy transport through 3- and 7-site models of the Fenna-Matthews-Olson photosynthetic light-harvesting complex, and the radical pair mechanism proposed for avian navigation under various rates of dephasing. For both of these systems, we demonstrate the ability to capture dynamics on a quantum simulator without loss of accuracy in the long-time limits.

This approach involves the vectorization of the Lindblad equation to retain the complete, and generally mixed, density matrix at each step of the system's evolution. This process has a quadratic overhead in the system dimension, doubling the size of the qubit space required for the simulation. While this is costly, this approach is in contrast to utilizing the operator-sum formulation, where knowledge of the time-dependent Kraus maps is required or additional approximations are necessary. The present approach does not rely on knowledge of the Kraus maps, avoids solving the differential equation on the original Hilbert space, and allows for direct simulation of the mixed state density matrix, albeit in unravelled form. When coupled with this Liouville space representation, the SVD-based algorithm allows for simulation of long-time dynamics in a way that requires only sparse, diagonal operations over the dilated $(k+1)$-qubit space, along with unitary operations on the original $k$-qubit space.  While other methods to encode non-unitary operators as unitary exist, such as the Sz.-Nagy dilation, they generally produce operators which act on the entirety of the dilated $(k+1)$-qubit space without inherent sparsity. In the present approach, after performing the classical singular value decomposition, the dilated non-unitary component that spans the $(k+1)$-qubit space is diagonal and can be implemented efficiently.~\cite{Welch:2014vj}

These results show progress towards using quantum algorithms to predict and explore quantum phenomena in biological processes; however, it should be noted that the systems studied are beyond the scope of possible implementation on current NISQ computers, with resource estimates discussed in the SI. The circuit complexity is dominated by implementation of the unitary evolution components. While there are likely cases where the SVD inherits exploitable symmetries from the original operators, $\hat{U}$ and $\hat{V}^{\dagger}$ may not retain this structure from  numerical calculation, resulting in dense operators in $k$-qubit space. Along with using structured or analytically available SVDs, techniques from unitary and Hamiltonian simulation could broaden the scope of systems that can be practically implemented on current NISQ hardware. Moreover, exploiting symmetries and structure in the operators to minimize circuit depth for this algorithm is an active area of on-going research; however, this challenge does not lessen the value of this approach. Notably, the SVD-based algorithm introduces no inherent limitation on the duration of a possible simulation, which is a challenge for several quantum dynamics algorithms. Here, we have demonstrated its success in capturing the dynamics of an exciton in a light-harvesting antenna and spins in a radical pair mechanism, showing its  promise for the accurate simulation of long-time dynamics for quantum biological systems. The efficacy of this algorithm could open up new pathways towards practical use of current quantum computers in predicting biologically-relevant quantum dynamics and steady states.

\section{Supporting Information}

Relevant parameter and variable tables for the Fenna-Matthews-Olson complex and avian compass models, more detailed descriptions of the operators and initial states for implementation,  and a brief quantum circuit resource estimate for a sample system (PDF). 

\section{Acknowledgements}

KHM acknowledges the start-up funds from Washington University in St. Louis.

\renewcommand*{\bibfont}{\normalsize}
\bibliography{main}

\end{document}